\documentclass[%
 reprint,
superscriptaddress,
amsmath,amssymb,
aps,
pra,
]{revtex4-1}
\usepackage{caption}
\usepackage{subcaption}
\usepackage{graphicx}
\usepackage{xcolor}
\usepackage{graphicx}
\usepackage{dcolumn}
\usepackage{bm}

\begin{document}

\preprint{APS/123-QED}

\title{Vector space-time wave packets}

\author{Murat Yessenov}
\affiliation{CREOL, The College of Optics \& Photonics, University of Central Florida, Orlando, FL 32816, USA}
\author{Zhaozhong Chen}
\affiliation{James Watt School of Engineering, University of Glasgow, UK}
\author{Martin P.~J. Lavery}
\affiliation{James Watt School of Engineering, University of Glasgow, UK}
\author{Ayman F. Abouraddy}
\thanks{corresponding author: raddy@creol.ucf.edu}
\affiliation{CREOL, The College of Optics \& Photonics, University of Central Florida, Orlando, FL 32816, USA}

\begin{abstract}
Space-time wave packets (STWPs) are propagation-invariant pulsed beams whose characteristics stem from the tight association between their spatial and temporal degrees of freedom. Until recently, only scalar STWPs have been synthesized in the form of light sheets. Here we synthesize vector STWPs that are localized in all dimensions by preparing polarization-structured spatio-temporal spectra and unveil the polarization distribution over the STWP volume via time-resolved complex field measurements. Such vector STWPs are endowed with cylindrically symmetric polarization vector structures, which require joint manipulation of the spatial, temporal, and polarization degrees-of-freedom of the optical field. These results may be useful in particle manipulation, and in nonlinear and quantum optics.
\end{abstract}



\maketitle

Bringing the various degrees of freedom (DoFs) of the optical field under \textit{joint} control helps expand the repertoire of accessible field structures, which can potentially open up vistas of new applications. For example, advances in sculpting the spatial distribution of the polarization vector has enabled the study of so-called `vector beams', which produce tighter focal spots \cite{Youngworth00OE,Qabis00OC,Dorn03PRL,Abouraddy06PRL,Bauer14NP}, among other possibilities \cite{Zhan09AOP,Eshaghi21SPIE}. Alternatively the temporal and polarization DoFs can be jointly manipulated to vary the polarization vector in time along an ultrashort pulse \cite{Brixner01OL,Brixner04OL}, for applications in quantum control \cite{Suzuki92PRL,Brixner04PRL} and nanoscale field structuring \cite{Aeschlimann07Nature}. More recently, space-time wave packets (STWPs) \cite{Yessenov19OPN,Yessenov22AOP} have been developed in which the spatial and temporal DoFs are jointly modulated to bring about propagation invariance \cite{Kondakci16OE,Parker16OE,Kondakci17NP}, tunable group velocity \cite{Salo01JOA,Wong17ACSP2,Kondakci19NC}, dispersion cancellation \cite{Yessenov21ACSP,Hall22LPR}, and novel device physics \cite{Shiri20NC,Shiri20OL,Guo21PRR}.

A new frontier for structured light aims at manipulating \textit{all} the field DoFs simultaneously: spatial, temporal, \textit{and} polarization. Recent efforts in this area include the study of topological polarization structures in space-time coupled optical fields \cite{Mounaix2020NC,Guo21Light,Shen21NC,Meng22OE}, and vector-beam polarization structures have been examined in the context of STWPs \cite{Diouf21OE}. However, until very recently, STWPs have been synthesized along only one transverse dimension (light sheets), so that the polar 2D transverse polarization distribution associated with vector beams is incompatible with the light-sheet geometry. Furthermore, time-resolved measurements to reveal the internal polarization structure of the STWP were not carried out in \cite{Diouf21OE}. New methodologies have emerged very recently for synthesizing STWPs that are localized in all dimensions \cite{Guo21Light,Pang21OL,Pang22OE,Yessenov22NC}, which may thus yield STWPs with cylindrical symmetry in both the field \textit{and} the polarization structure.

\begin{figure}[b!]
\centering
\includegraphics[width=6cm]{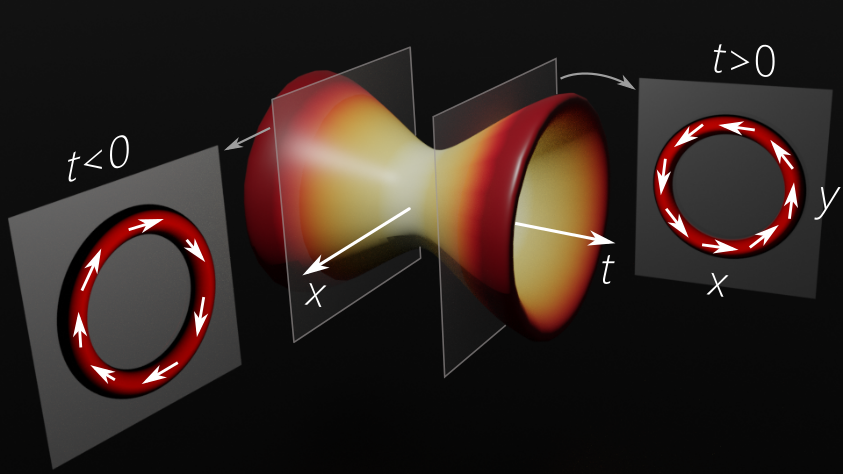}
\caption{Structure of a vector STWP with azimuthal polarization symmetry at a fixed axial plane. We plot the spatio-temporal intensity profile $I(x,y;t)$ at a fixed $z$ and superimpose the distribution of the polarization vector at $t\!>\!0$ and $t\!<\!0$.}
\label{Fig:ConceptWavePacketStructure}
\end{figure}

Here we synthesize and characterize \textit{vector} STWPs: propagation-invariant, space-time coupled pulsed beams that are localized in all dimensions, and simultaneously feature a spatio-temporally structured polarization vector, as illustrated in Fig.~\ref{Fig:ConceptWavePacketStructure}. Such field structures necessitate simultaneously modulating both transverse spatial dimensions, time, and the polarization vector to produce cylindrically symmetric vector light bullets that are propagation invariant \textit{and} localized. We realize vector STWPs with radial and azimuthal polarization symmetry and confirm their propagation invariance. Time-resolved measurements of the complex field distribution unveil the polarization structure of the 3D vector STWP field. By combining these unique characteristics with the demonstrated exquisite control over the group velocity for STWPs \cite{Kondakci19NC} and the possibility of incorporating orbital angular momentum \cite{Pang22OE,Yessenov22NC}, we expect that vector STWPs can enable the exploration of new phase-matching conditions in nonlinear optical processes.

\begin{figure}[t!]
\centering
\includegraphics[width=8.6cm]{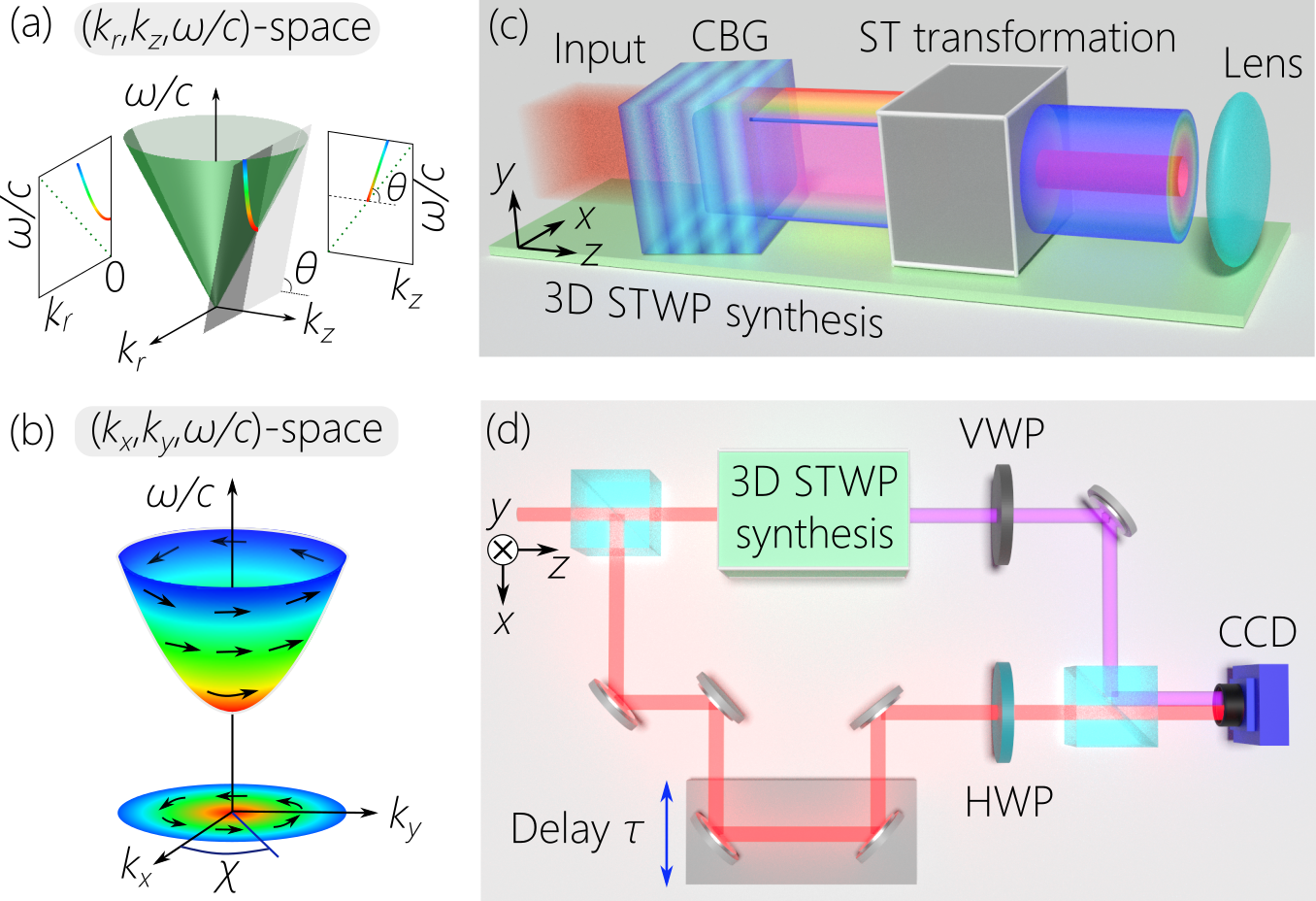}
\caption{(a) The spectral support for a superluminal 3D STWP on the surface of the free-space light-cone $k_{x}^{2}+k_{y}^{2}+k_{z}^{2}\!=\!(\tfrac{\omega}{c})^{2}$ and its spectral projections in $(k_{r},k_{z},\tfrac{\omega}{c})$-space, and (b) in $(k_{x},k_{y},\tfrac{\omega}{c})$-space. (c) Schematic of the optical setup for synthesizing 3D ST wave packets, and (d) for interferometrically characterizing 3D ST wave packets. STWP: Space-time wave packet; VWP: vortex wave plate; HWP: half-wave plate; CBG: volume chirped Bragg grating.}
\label{Fig:Setup}\vspace{-3mm}
\end{figure}

We first formulate vector STWPs theoretically. Propagation invariance of STWPs necessitates introducing a one-to-one spectral association between the transverse radial wave number $k_{r}$ and the temporal frequency $\omega$ \cite{Yessenov22AOP}: $\tfrac{\Omega}{\omega_{\mathrm{o}}}\!\approx\!\tfrac{k_{r}^{2}}{2k_{\mathrm{o}}^{2}(1-\cot{\theta})}$ in the paraxial regime; here $\omega_{\mathrm{o}}$ is a carrier frequency, $\Omega\!=\!\omega-\omega_{\mathrm{o}}$, $k_{\mathrm{o}}\!=\!\tfrac{\omega_{\mathrm{o}}}{c}$ is the associated wave number, $c$ is the speed of light in vacuum, and the physical significance of the angle $\theta$ will become clear shortly. This particular form of radial angular dispersion ensures that the axial wave number $k_{z}$ is linearly related to $\omega$: $k_{z}\!=\!k_{\mathrm{o}}+\Omega/\widetilde{v}$, where $\widetilde{v}\!=\!c\tan{\theta}$ \cite{Kondakci17NP,Yessenov22NC}. This is the equation for a plane in $(k_{r},k_{z},\tfrac{\omega}{c})$-space that is parallel to the $k_{r}$ axis and makes an angle $\theta$ (the `spectral tilt angle') with the $k_{z}$-axis [Fig.~\ref{Fig:Setup}(a)]. Such a plane intersects with the surface of the light-cone $k_{r}^{2}+k_{z}^{2}\!=\!(\tfrac{\omega}{c})^{2}$ in a conic section: an ellipse when $|\widetilde{v}|\!<\!c$ and a hyperbola when $|\widetilde{v}|\!>\!c$, both of which are approximated by a parabola in the paraxial regime where $k_{r}\!\ll\!k_{\mathrm{o}}$. In this construction, the envelope $\psi$ of a scalar STWP, $E(r,z;t)\!=\!e^{i(k_{\mathrm{o}}z-\omega_{\mathrm{o}}t)}\psi(r,z;t)$, travels rigidly in free space (without diffraction or dispersion) at a group velocity $\widetilde{v}$: $\psi(r,z;t)\!=\!\psi(r,0;t-z/\widetilde{v})$ \cite{Kondakci19NC,Yessenov22NC}.

When represented in terms of the Cartesian components in $(k_{x},k_{y},\tfrac{\omega}{c})$-space, the spectral support for a STWP is a conic surface of revolution (an ellipsoid or hyperboloid), as shown in Fig.~\ref{Fig:Setup}(b). Each frequency $\omega$ is associated with a circle comprising all wave vectors whose transverse components $k_{x}\!=\!\tfrac{\omega}{c}\sin{\varphi}\cos{\chi}$ and $k_{y}\!=\!\tfrac{\omega}{c}\sin{\varphi}\sin{\chi}$ have the same radial magnitude $k_{r}\!=\!\sqrt{k_{x}^{2}+k_{y}^{2}}\!=\!\tfrac{\omega}{c}\sin{\varphi}$. We assign to \textit{each} wave vector a different polarization unit vector; e.g., an \textit{azimuthally} symmetric polarization field $\hat{e}_{\mathrm{a}}(\chi)\!=\!-(\sin{\chi})\hat{x}+(\cos{\chi})\hat{y}$ [Fig.~\ref{Fig:Setup}(b)], or a \textit{radially} symmetric polarization field $\hat{e}_{\mathrm{r}}(\chi)\!=\!(\cos{\chi})\hat{x}+(\sin{\chi})\hat{y}$. In both cases the polarization vector is independent of the radial coordinate (but vanishes at the origin). We show below that this polarization-structured spatio-temporal spectrum is converted in physical space to a 3D STWP with the same polarization structure distributed in space and time.

To synthesize the 3D STWPs, we do \textit{not} follow the approach developed in \cite{Kondakci17NP,Kondakci19NC} that involves using a spatial light modulator (SLM) to impart a 2D phase distribution to the spectrally resolved wave front. As pointed out in \cite{Wefers96OL}, this type of approach can only modulate the field along one spatial dimension, because the other SLM dimension is reserved for modulating the temporal spectrum. Therefore, we make use instead of the technique introduced in \cite{Yessenov22NC} that engenders a prescribed \textit{radial} angular dispersion profile encompassing both transverse dimensions via a three-stage experimental strategy [Fig.~\ref{Fig:Setup}(c)]. First, starting with generic plane-wave femtosecond pulses (Tsunami, Spectra Physics; $\sim\!100$-fs pulses centered at a wavelength of $\approx\!796$~nm), a volume chirped Bragg grating (CBG; OptiGrate L1-02) in a double-pass configuration spreads the pulse spectrum along one transverse dimension while maintaining a flat phase front. In contrast, the spectrally resolved wave front produced by a conventional diffraction grating has a rapidly varying phase modulation. The 34-mm-long CBG has a central axial periodicity of 270~nm, a chirp rate of $-30$~pm/mm, and an average refractive index of $n\!=\!1.5$, resulting in a  linear spatial chirp of $\approx\!-22.2$~mm/nm at $\lambda_{\mathrm{o}}\!=\!796$~nm \cite{Kaim14SPIE,Glebov14SPIE} over a bandwidth of $\approx\!0.3$~nm.

Second, a spectral transformation implemented via a pair of SLMs \cite{Bryngdahl74JOSA} reshuffles the wavelengths along one dimension in space to produce an arbitrary linear chirp as required to tune the group velocity of the STWP \cite{Yessenov22NC}. Third, a log-polar geometrical transformation is implemented using a sequence of two axially separated phase distributions, whereby each line associated with a single wavelength in the incident field is transformed at the output into a circle \cite{Bryngdahl74JOSA,Hossack87JOMO} (i.e., operated in the opposite direction as that implemented in \cite{Berkhout10PRL,Lavery12OE}). The result is a spectrally resolved wave front in which the wavelengths are arranged radially as a desired sequence of concentric circles, corresponding to the spectral support in Fig.~\ref{Fig:Setup}(b). The second and third stages are combined in Fig.~\ref{Fig:Setup}(c) within the spatio-temporal (ST) transformation. Finally, a Fourier transforming spherical lens (300-mm focal length) produces the 3D STWP in physical space [Fig.~\ref{Fig:Setup}(c)]. The resulting structure of the scalar propagation-invariant STWP at a fixed axial plane $z$ is rotationally symmetric around the propagation axis, and the spatio-temporal profile is X-shaped in any meridional plane. We synthesize radially and azimuthally polarized vector STWPs travelling at a superluminal group velocity $\widetilde{v}\!\!=\!\!1.2c$ ($\theta\!=\!50^{\circ}$). 

\begin{figure}[t!]
\centering
\includegraphics[width=8.6cm]{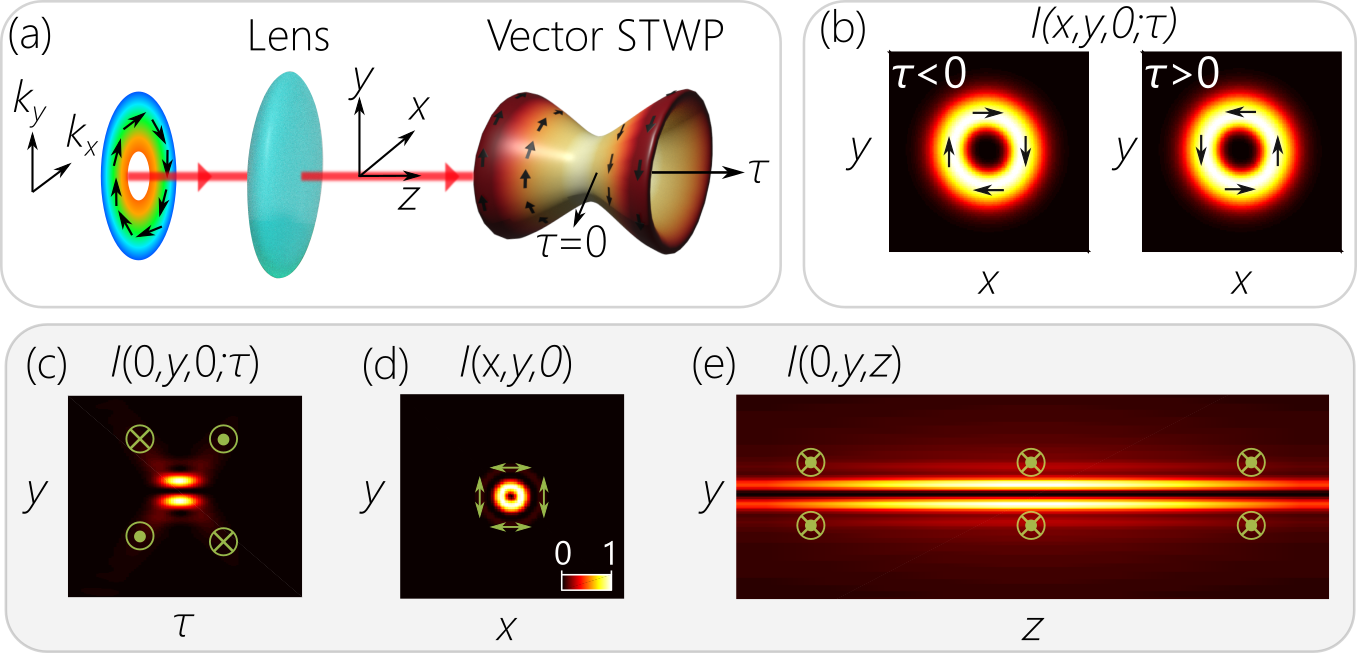}
\caption{(a) Conceptual illustration of the formation of a vector STWP with azimuthally symmetric polarization plotted at a fixed axial plane after the Fourier transform of the polarization-modulated spatio-temporal spectrum. (b) The vector STWP intensity $I(x,y,0;\tau)=|E(x,y,z\!=\!0;\tau)|^{2}$ at a fixed axial plane at two different $\tau$: $\tau\!<\!0$ and $\tau\!>\!0$. (c) The vector STWP intensity $I(0,y,0;\tau)$ at a fixed axial plane $z$, in a meridional plane $x\!=\!0$. (d) The time-averaged intensity $I(x,y,0)\!=\!\int\!d\tau\,I(x,y,0;\tau)$ at a fixed axial plane, and (e) $I(0,y,z)$ along $z$ in a meridional plane. In (b-e), we overlay the polarization-vector distribution on the intensity.}
\label{Fig:Concept}
\end{figure}

To introduce the structured polarization into the 3D STWP, we place a vortex wave plate (VWP, Thorlabs WPV10L-780) immediately after the lens [Fig.~\ref{Fig:Setup}(d)]. It is conceptually clearer to place the VWP in the plane immediately preceding the lens [Fig.~\ref{Fig:Concept}(a)], whereby each point in this spectral plane corresponds to a single wavelength $\lambda$ \textit{and} transverse wave number $k_{r}(\lambda)$, and the VWP changes the polarization along the azimuthal angle $\chi$. Several changes occur in the STWP structure in presence of the azimuthally or radially symmetric polarization vector field. First, a null is formed in the intensity along the propagation axis [Fig.~\ref{Fig:ConceptWavePacketStructure}]. This is clear whether the time-resolved intensity is examined at a fixed plane $z$, $I(x,y,0;\tau)\!=\!|E(x,y,z\!=\!0;\tau)|^{2}$ [Fig.~\ref{Fig:Concept}(b,c)], or in the time-averaged intensity $I(x,y,z)\!=\!\int\!d\tau\,I(x,y,z;\tau)$ [Fig.~\ref{Fig:Concept}(d,e)]. Second, at a fixed axial plane $z$, the polarization vector is distributed over the 3D spatio-temporal structure of the STWP. At each time $\tau$, the field in the $(x,y)$-plane is endowed with the vector beam structure encoded at the source in the spectral plane [Fig.~\ref{Fig:Concept}(b,c)].

\begin{figure}[t!]
\centering
\includegraphics[width=8.6cm]{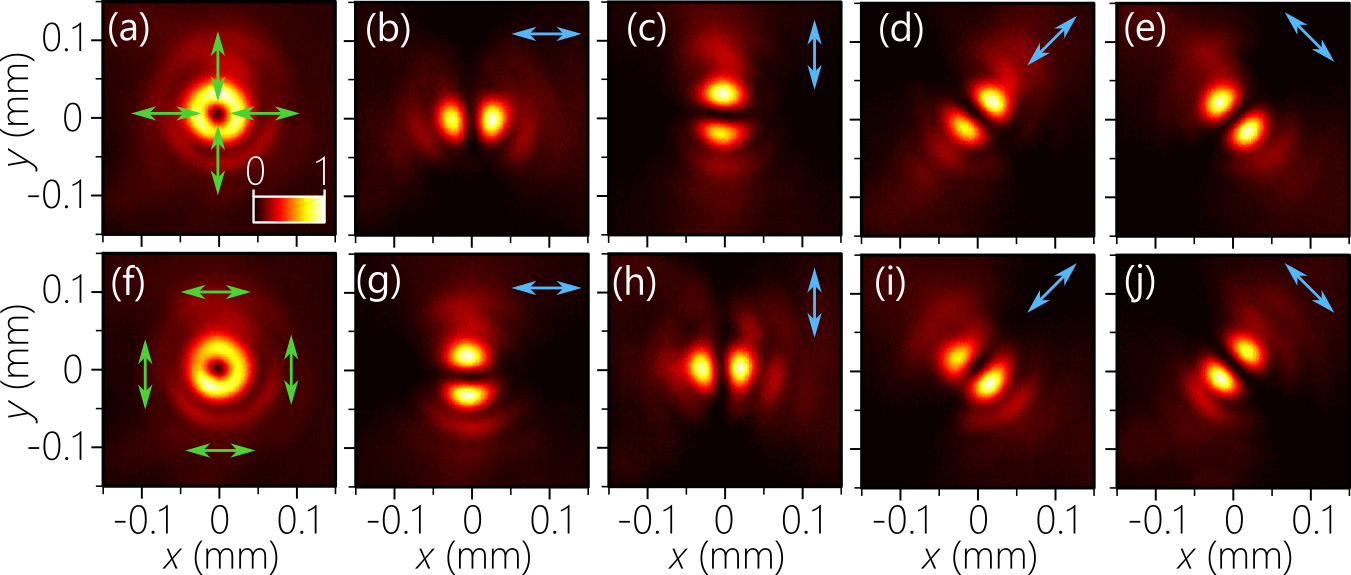}
\caption{Intensity profiles $I(x,y)$ of vector STWPs at a fixed axial plane $z\!=\!50$~mm with (a-e) radial or (f-j) azimuthal polarization symmetry. (a,f) Total intensity distribution. The remaining panels are the intensity profiles projected onto the polarization indicated by the arrow in the top right corner.}
\label{Fig:TimeAveragedProjections} \vspace{-5mm}
\end{figure}

Third, the overall direction of the polarization at each position in the $(x,y)$-plane in the time-resolved field structure is reversed at $\tau\!>\!0$ with respect to that at $\tau\!<\!0$ [Fig.~\ref{Fig:Concept}(b,c)]: $\hat{e}_{\mathrm{a}}(\chi)\!\rightarrow\!-\hat{e}_{\mathrm{a}}(\chi)$. A similar inversion occurs for the radially symmetric polarization configuration $\hat{e}_{\mathrm{r}}(\chi)\!\rightarrow\!-\hat{e}_{\mathrm{r}}(\chi)$. This can be understood as a manifestation of `time-diffraction' \cite{Longhi04OE,Porras17OL} whereby the structure at fixed $z$ of the non-diffracting STWP $E_{\mathrm{STWP}}(x,y,0;\tau)$ corresponds to the structure of a monochromatic beam $E_{\mathrm{mono}}(x,y,z)$ after replacing $z$ in $E_{\mathrm{mono}}$ with an appropriately scaled $\tau$ in $E_{\mathrm{STWP}}$ when initially $E_{\mathrm{STWP}}(x,y,0;0)\!=\!E_{\mathrm{mono}}(x,y,0)$. In other words, the evolution of the monochromatic beam along $z$ (diffractive spreading) is displayed by the STWP in time $\tau$ at fixed $z$. This is particularly clear in the vector STWP field structure in a meriodional plane [Fig.~\ref{Fig:Concept}(c)]. Fourth, the vector STWP is propagation invariant like its scalar counterpart [Fig.~\ref{Fig:Concept}(e)]. Note the double-sided arrows at each position in the time-averaged intensity [Fig.~\ref{Fig:Concept}(d,e)] resulting from integration over time (whereupon the polarization vector changes by a phase $\pi$) at each position. This distinguishes the vector STWP from its monochromatic counterpart, where the vector is fully defined at each point.

\begin{figure}[t!]
\centering
\includegraphics[width=8.6cm]{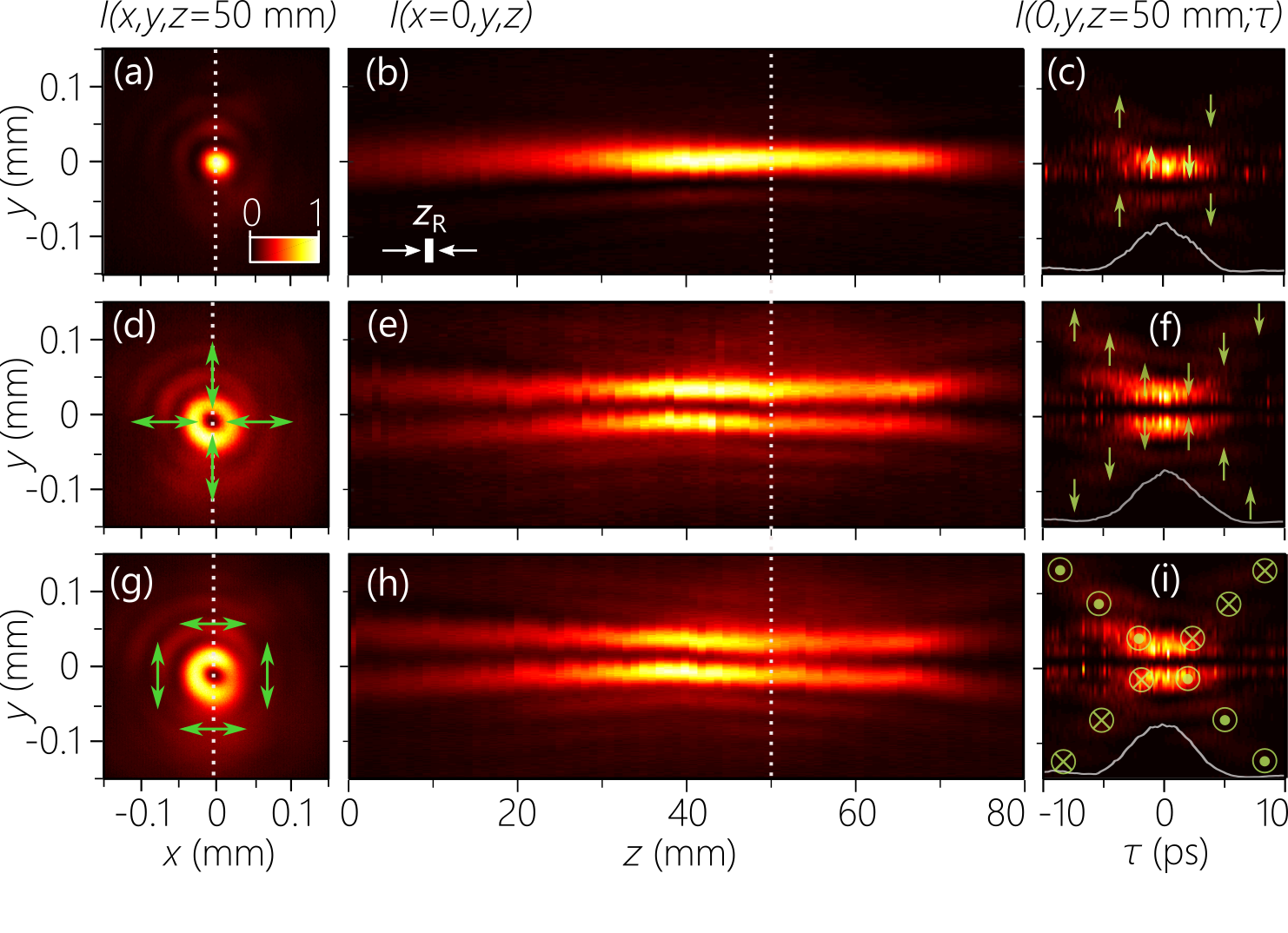}
\caption{(a) Time-averaged intensity $I(x,y,z\!=\!50~\mathrm{mm})$ for a linearly polarized STWP, (b) the axial evolution of the time-averaged intensity in a meridional plane $I(x\!=\!0,y,z)$, and (c) the time-resolved intensity profile $I(x\!=\!0,y,z\!=\!50~\mathrm{mm};\tau)$. The white bar in (b) corresponds to the Rayleigh range of a Gaussian beam having the same transverse width of the STWP in (a). (d-f) Same as (a-c) for a radially polarized vector STWP. (g-i) Same as (a-c) for an azimuthally polarized vector STWP.}
\label{Fig:PropInvAndTimeResolvedProfile}\vspace{-3mm}
\end{figure}

\begin{figure}[t!]
\centering
\includegraphics[width=8.6cm]{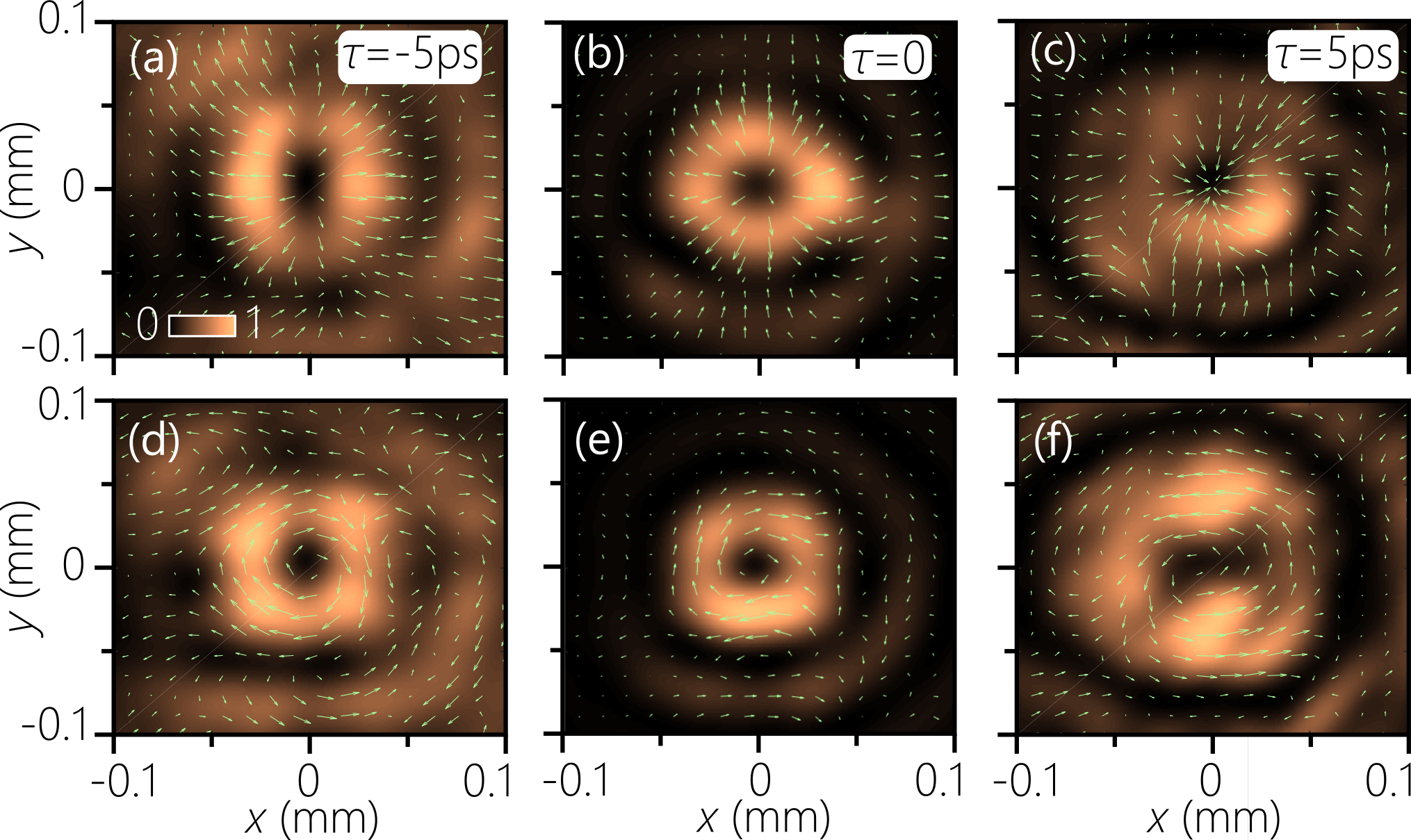}
\caption{Reconstructed polarization structure $\mathbf{E}(x,y,z\!=\!50~\mathrm{mm};\tau)$ at $\tau\!=\!-5,0,5$~ps for (a-c) radially polarized and (d-f) azimuthally polarized vector STWPs. To improve the visualization, we plot as a background the total intensity $I(x,y,z\!=\!50~\mathrm{mm},\tau)$. The polarization is represented by green arrows whose direction and length correspond to those of the vector field $\mathbf{E}(x,y,z\!=\!50~\mathrm{mm};\tau)$.}
\label{Fig:VectorWP}\vspace{-3mm}
\end{figure}

In our experiments we measure along the different polarization axes the time-averaged intensity $I(x,y,z)$ [Fig.~\ref{Fig:TimeAveragedProjections} and Fig.~\ref{Fig:PropInvAndTimeResolvedProfile}], the time-resolved intensity $I(x,y,z;\tau)$ [Fig.~\ref{Fig:PropInvAndTimeResolvedProfile}], and the time-resolved complex field $\mathbf{E}(x,y,z;\tau)$ [Fig.~\ref{Fig:VectorWP}], following the methodology developed in \cite{Yessenov22NC}. First, we acquire polarization projections of the time-averaged transverse intensity profiles at a fixed axial position $z\!=\!50$~mm by placing a CCD camera (TheImagingSource, DMK27BUP031) and polarizer in the path of the STWP [Fig.~\ref{Fig:TimeAveragedProjections}]. We plot the measurement results for a radially polarized vector STWP in Fig.~\ref{Fig:TimeAveragedProjections}(a-e), and for an azimuthally polarized vector STWP in Fig.~\ref{Fig:TimeAveragedProjections}(f-j). The intensity profiles at $z\!=\!50$~mm with no polarizer are donut-shaped with a radius of $\approx\!35~\mu$m at the peak intensity [Fig.~\ref{Fig:TimeAveragedProjections}(a,f)], which become the typical Hermite-Gaussian-like field structure after the linear polarizer. Moreover, we verify the propagation invariance of the vector STWPs by scanning the CCD camera along $z$-axis to record the axial evolution of the time-averaged intensity. We carry out this measurement for a scalar STWP as a reference in Fig.~\ref{Fig:PropInvAndTimeResolvedProfile}(b), where diffraction-free propagation over $L\!\approx\!40$~mm is observed with a central lobe having FWHM$\approx\!30~\mu$m. Similar diffraction-free behavior is observed for the radially and azimuthally polarized vector STWPs [Fig.~\ref{Fig:PropInvAndTimeResolvedProfile}(e,h)].

The time-resolved intensity profile $I(x,y,z;\tau)$ at a fixed axial plane $z$ is reconstructed by placing the vector STWP synthesis setup in one arm of an interferometer [Fig.~\ref{Fig:Setup}(d)] while utilizing the linearly polarized (tuned via a half-wave plate [Fig.~\ref{Fig:Setup}(d)]) short input plane-wave pulse as a reference in the second arm. By sweeping an optical delay line in the reference path, the visibility of the interferogram recorded when the reference pulse and STWP overlap in space and time helps reconstruct the spatio-temporal intensity profiles $I(x,y,z;\tau)$. We plot in Fig.~\ref{Fig:PropInvAndTimeResolvedProfile}(c,f,i) the reconstructed profiles for the scalar, radially polarized, and azimuthally polarized STWPs in a meridional plane $x\!=\!0$. These time-resolved measurements reveal clearly the expected X-shaped profile with a peak at $y\!=\!0$ for the scalar STWP [Fig.~\ref{Fig:PropInvAndTimeResolvedProfile}(c)] and a dip at the center for the vector STWP counterparts [Fig.~\ref{Fig:PropInvAndTimeResolvedProfile}(f,i)]. The radially and azimuthally polarized vector STWPs have different polarization states in the meridional plane $x\!=\!0$, which corresponds to the polarization of the reference pulse that interferes with the vector STWP.

Simulations predict that the polarization vector switches sign upon passing through $\tau$, but the time-resolved intensity measurements in Fig.~\ref{Fig:PropInvAndTimeResolvedProfile} cannot directly confirm that behavior. Instead, we add a small angle between propagation directions of the reference pulse and STWP [Fig.~\ref{Fig:Setup}(d)] to produce off-axis interference fringes from which we reconstruct the complex field distribution for each polarization component via the off-axis digital-holography algorithm \cite{Cuche99OL,Sanchez-Ortiga14AO} (see \cite{Yessenov22NC} for more details), and hence the complex field $\mathbf{E}(x,y,z;\tau)$. The field is recorded at three points in time $\tau\!=\!-5,0,5$~ps at the fixed axial position $z\!=\!50$~mm. The expected topological polarization structure is confirmed for the radially polarized STWP is revealed, including the polarization flips occuring in the side lobes [Fig.~\ref{Fig:VectorWP}(a-c)]. Moreover, the measurements confirm the predicted flip in polarization direction at each point in space for $\tau\!>\!0$ with respect to that for $\tau\!<\!0$. In Fig.~\ref{Fig:VectorWP}(d-f), similar behaviour is observed for the azimuthally polarized vector STWP.

In conclusion, we have synthesized and characterized propagation-invariant vector STWPs in which the spatial, temporal, and polarization degrees of freedom are inextricably intertwined. The STWPs are localized along all dimensions, and are endowed with a structured polarization distribution over its spatio-temporal volume, which is maintained for extended propagation distances. These results may be useful for novel experiments in particle control or optical communications through turbulent environments. Moreover, it has been argued that combining control over the orbital angular momentum (as demonstrated in \cite{Yessenov22NC}) and the polarization structure (as demonstrated here) can impact the photon statistics of quantum states of light \cite{Yang21CP}. Finally, one may also consider combining this approach with recent techniques utilizing metasurfaces to also vary the polarization state along the propagation axis \cite{Dorrah21NP}.
\\ \\
\textbf{Funding.}
U.S. Office of Naval Research (ONR) N00014-17-1-2458 and N00014-20-1-2789.
\\ \\
\textbf{Disclosures.}
The authors declare no conflicts of interest.
\\ \\
\textbf{Data availability.}
Data underlying the results presented in this paper are not publicly available at this time but may be obtained from the authors upon reasonable request.
\clearpage
\bibliography{diffraction}

\end{document}